\documentclass[10pt,sn-nature]{sn-jnl}
\usepackage{amsmath,amssymb,amsfonts,stmaryrd,wasysym,graphicx,multirow,color,textcomp,subfigure}
\usepackage{manyfoot}
\usepackage{float}
\usepackage{color}
\usepackage{soul}
\usepackage{bm}
\usepackage[normalem]{ulem}
\usepackage[dvipsnames]{xcolor}
\usepackage{geometry}
\setcitestyle{super,open={},close={}}

\renewcommand{\v}[1]{{\bf #1}}

\begin{document}

\title{Scalar Spin Chirality Hall Effect}

\author[1]{\fnm{Gyungchoon} \sur{Go}}
\author[1]{\fnm{Durga Prasad} \sur{Goli}}
\author[2,3]{\fnm{Nanse} \sur{Esaki}}
\author[4]{\fnm{Yaroslav} \sur{Tserkovnyak}}

\author*[1]{\fnm{Se Kwon} \sur{Kim}} \email{sekwonkim@kaist.ac.kr}

\affil*[1]{Department of Physics, Korea Advanced Institute of Science and Technology, Daejeon 34141, Korea}
\affil[2]{Department of Physics, Graduate School of Science, The University of Tokyo, 7-3-1 Hongo, Tokyo 113-0033, Japan}
\affil[3]{Advanced Science Research Center, Japan Atomic Energy Agency, Tokai, Ibaraki 319-1195, Japan}
\affil[4]{Department of Physics and Astronomy and Bhaumik Institute for Theoretical Physics, University of California, Los Angeles, Los Angeles, CA, 90095, USA}

\abstract{
The scalar spin chirality, which characterizes the fundamental unit of noncoplanar spin structures, plays an important role in rich chiral physics of magnetic materials. In particular, the intensive research efforts over the past two decades have demonstrated that the scalar spin chirality is the source of various novel Hall transports in solid-state systems, offering a primary 	route to bring about chiral phenomena in condensed matter physics. However, in all of the previous studies, the scalar spin chirality has been given as a stationary background, serving only a passive role in the transport properties of materials. It remains an open question whether or not the scalar spin chirality itself can exhibit a Hall-type transport. In this work, we show that the answer is yes: The scalar spin chirality is Hall-transported in Kagome ferromagnets and antiferromagnets under an external bias, engendering a phenomenon which we dub the scalar spin chirality Hall effect. Notably, this effect is present even in the absence of any spin-orbit coupling. The analytical theory for the scalar spin chirality Hall effect is corroborated by atomistic spin simulations. Our findings call for the need to lift the conventional assumption that the scalar spin chirality is a passive background in order to discover the active roles of the scalar spin chirality in transport properties.
}

\maketitle


\section{Introduction}

\begin{figure}[t]
\centering
\includegraphics[width=0.5\columnwidth]{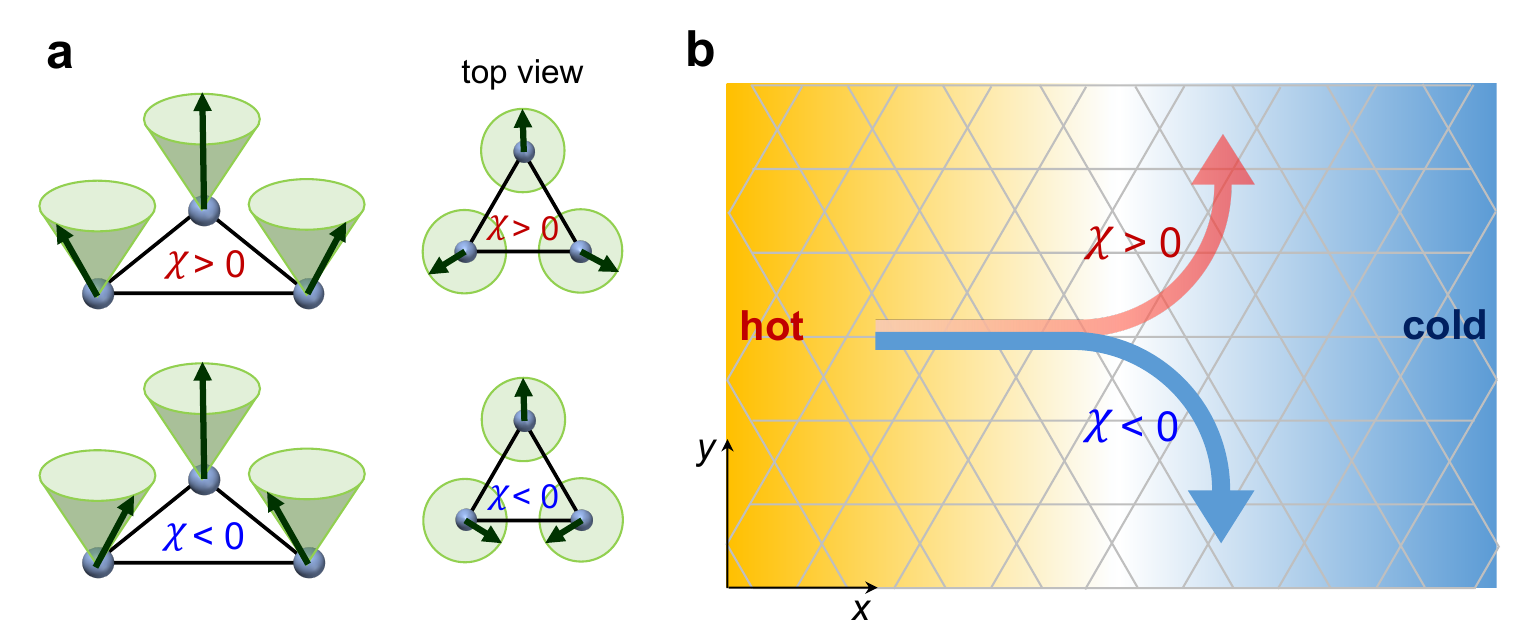}
\caption{Schematic illustration of \textbf{a}, the spin configurations with positive SSC ($\chi > 0$) and negative SSC ($\chi < 0$), and \textbf{b}, the SSC Nernst effect.}\label{fig:1}
\end{figure}

A triple product of spins $\chi = \v {\hat S}_i \cdot ({\v {\hat S}_j \times \v {\hat S}_k})$,
known as the scalar spin chirality (SSC) (see Fig.~\ref{fig:1}\textbf{a}), is increasingly recognized as a key concept for understanding magnetic systems. It characterizes the most elementary unit of noncoplanar spin structures and plays a crucial role in understanding rich chiral physics of magnetic materials.
In quantum magnetism, it has been introduced to describe the chiral spin state which breaks time-reversal symmetry in relation to superconductivity~\cite{Wen1989}.
Also, the SSC has been found to contribute to the orbital magnetization through electron hopping among triplets of noncoplanar spins~\cite{Hoffmann2015b, Dias2016, Hanke2016a, Lux2018a, Grytsiuk2020, Zhang2020a}. In the transport properties of solid-state systems, the SSC is known to induce an effective magnetic field through the Berry curvature and also serve as a scattering source on particles in solids such as conduction electrons~\cite{Taguchi2001,Tatara2002,Machida2010,Ishizuka2018}, magnons~\cite{Katsura2010}, and phonons~\cite{Oh2024}. Consequently, it leads to the Hall-like transports of the aforementioned particles in the frustrated magnets~\cite{Taguchi2001,Machida2010,Shindou2001,Nakatsuji2015,Katsura2010,Han2017,Kim2024} and
in chiral magnets~\cite{Neubauer2009,Kanazawa2011,Franz2014,Ishizuka2018,vanHoogdalem2013,Kim2019b,Xu2024} through both intrinsic and skew scattering mechanisms. It is worth noting that, in previous studies, the SSC has been given as an immobile background that passively influences the dynamics of the particles in solids.
The transport properties of SSCs themselves have not been discussed before, in stark contrast to the previous extensive studies on the transport of macroscopic chiral spin textures such as magnetic skyrmions. In particular, it is an open question whether the SSC can exhibit a Hall-like transport when it is allowed to be dynamic, which we would like to address in this work. From the symmetry point of view, SSC, $\chi = \v {\hat S}_i \cdot ({\v {\hat S}_j \times \v {\hat S}_k})$, possesses the same symmetry properties as $B_z$ or $S_z$ with $z$-axis being normal to the triangular plane defined the three sites, since it remains scalar under proper or improper spin rotations and transforms as the $z$-component of a magnetic field under spatial transformation. Since $S_z$ is known to exhibit Hall transports via the spin-orbit coupling, e.g., the spin Hall effect and the spin Nernst effect, it is natural to ask whether the SSC with the same symmetry property can exhibit a Hall transport, and, by going one step further, if it can, whether the SSC Hall effect can be present without any spin-orbit coupling.

Here, we theoretically demonstrate that the SSC exhibits an intrinsic Hall transport in Kagome ferromagnets and Kagome antiferromagnets within the theoretical framework where the SSC is described in terms of the magnon operators. We show that the external biases, such as a temperature gradient or a magnon-current injection, induce a SSC flow,
perpendicular to the direction of the external perturbation. We refer to this phenomenon as the SSC Hall effect.
More precisely, nonequilibrium magnonic states with positive and negative SSCs exhibit opposite Hall transports (Fig.~\ref{fig:1}\textbf{b}).
Interestingly, unlike magnon Hall-like phenomena explored in the previous studies, such as the magnon Hall effect~\cite{Katsura2010,Onose2010,Matsumoto2011,Matsumoto2014a,Mook2014a,Kim2016a,Owerre2016}, the magnon spin Nernst effect~\cite{Cheng2016,Zyuzin2016,Shiomi2017,Zhang2022},
and the magnon orbital Nernst effect~\cite{Go2024b}, the SSC Hall effect does not rely on the magnon Berry curvature and therefore does not require spin-orbit coupling;
even if the magnon Berry curvature is zero everywhere, the SSC Hall effect still exists.
Thus, the SSC Hall effect is an inherent property of the Kagome magnets without requiring any specific symmetry breaking or spin-orbit-coupling related effects.
We further verify the SSC Hall effect through computer experiments by simulating the atomistic spin dynamics in the Kagome magnets. We envision that our findings trigger further investigations of novel transport phenomena where the SSC plays active roles, enriching the chiral physics of solid-state systems.

\section{Main results}

\begin{figure}[t]
\includegraphics[width=1.0\columnwidth]{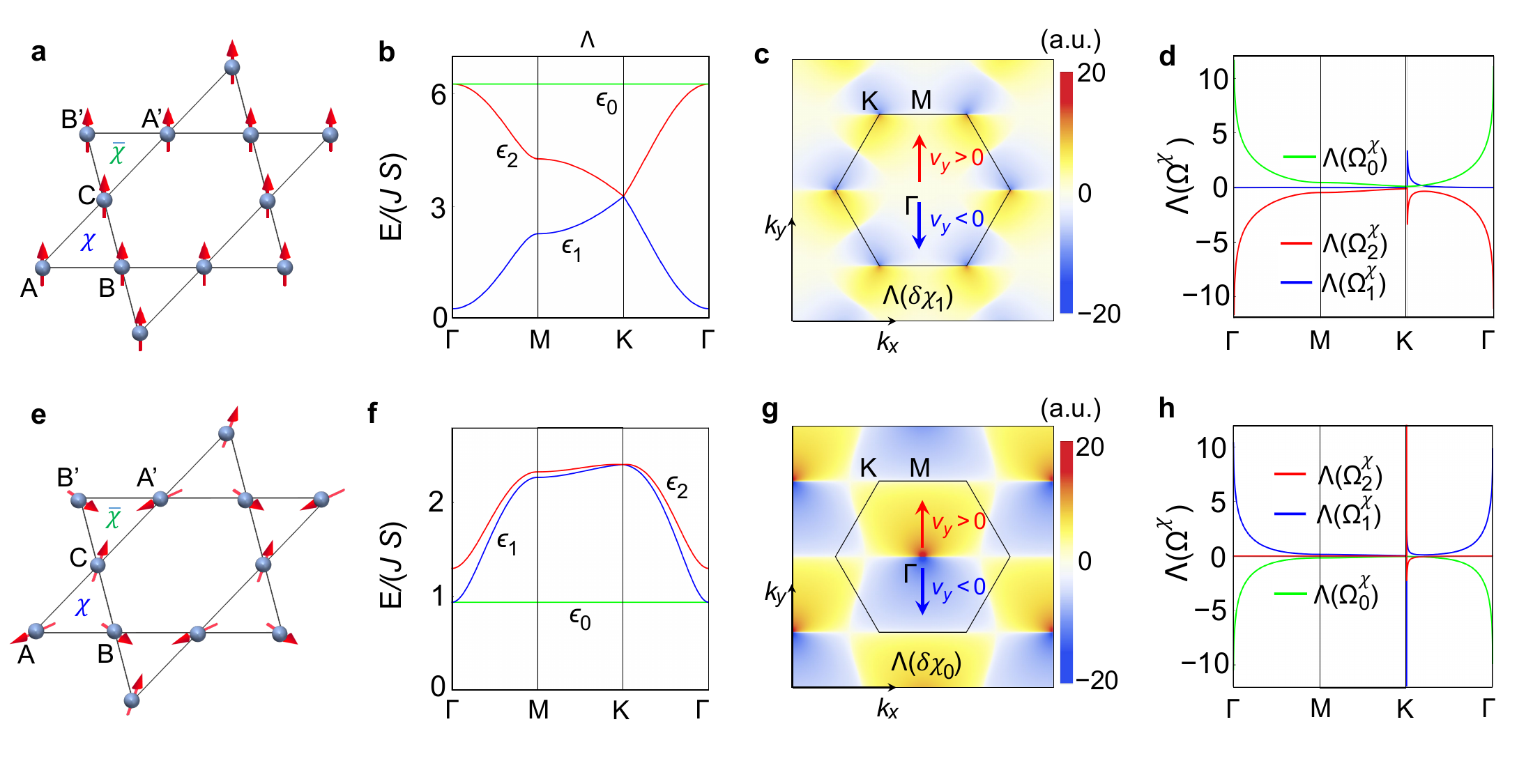}
\caption{\textbf{a}, Ground state spin configuration, \textbf{b}, magnon dispersion, \textbf{c}, nonequilibrium SSC of the lowest magnonic state $\delta \chi_1$ in a log-scale
[$\Lambda(\delta\chi_1) \equiv {\rm sgn}(\delta\chi_1) {\rm log} (1 + |\delta\chi_1|)$] in the presence of temperature gradient along $x$-direction, and, \textbf{d}, log-scale SSC Berry curvature
[$\Lambda(\Omega^\chi) \equiv {\rm sgn}(\Omega^\chi) {\rm log} (1 + |\Omega^\chi|)$] of the collinear Kagome FM.
\textbf{e}, Ground state spin configuration, \textbf{f}, magnon dispersion, \textbf{g}, nonequilibrium SSC of the lowest magnon state $\delta \chi_0$ in a log-scale in the presence of temperature gradient along $x$-direction, and, \textbf{h}, log-scale SSC Berry curvature of the coplanar Kagome AFM.}\label{fig:2}
\end{figure}

Here, we consider the following spin Hamiltonian in the Kagome lattice
\begin{align}\label{H0}
H = -J \sum_{\langle i,j \rangle} \v S_i\cdot \v S_j - K \sum_i ({\v S_i\cdot {\v e}_i})^2 \, ,
\end{align}
where $J$ is the exchange coupling and $K$ is the easy-axis magnetic anisotropy along ${\v e}_i$.
In this study, we explore two distinct magnetic systems: the collinear Kagome ferromagnet (FM) ($J >0$, $K>0$, and a uniform easy-axis, e.g., ${\v e}_i = \hat{\v z}$) and the
coplanar Kagome antiferromagnet (AFM) with 120$^\circ$ spin order ($J < 0$ and $K > 0$). In the coplanar Kagome AFM model, the easy-axis directions are sublattice dependent and collinear to the spin directions shown in Fig.~\ref{fig:2}\textbf{e}.
In the main text, we focus on the Hamiltonian of Eq.~\eqref{H0} without considering the external magnetic field and the Dzyaloshinskii-Moriya interaction (DMI),
resulting in vanishing magnon Berry curvatures across the entire momentum space.
The effects of the external magnetic field and the DMI are discussed in the Supplementary Information.

The ground states of both FM and AFM systems possess no SSC. Our interest is in the thermally excited SSC on top of these ground states and their Hall transports. Since elementary excitations of the considered magnetic systems are magnons, it is natural to describe the SSC in terms of the magnon operators. By employing the Holstein-Primakoff approach, we obtain the magnon dispersions of the FM and the AFM models, which are shown in Fig.~\ref{fig:2}\textbf{b} and \textbf{f}, respectively.
For the collinear FM, we use the model parameter of V$_3$Br$_8$Li: $J = 1.13$ meV, $K = 0.152$ meV, $S = 5/2$~\cite{You2023}.
For the coplanar AFM, we simply flip the sign of the exchange constant ($J = -1.13$ meV), while keeping the same values of $K$ and $S$ as in the collinear FM.
In both models, the magnon spectra consist of two dispersive bands ($\epsilon_1$ and $\epsilon_2$) and a flat band $\epsilon_0$, with no band gap between the different bands.
For the Kagome FM, the flat band is located at the top of the band dispersion $[\epsilon^{\rm FM}_0 = \epsilon^{\rm FM}_2({\Gamma}) = 6 J S + 2K S]$,
while for the coplanar AFM it is located at the bottom of the band dispersion $[\epsilon^{\rm AFM}_0 = \epsilon^{\rm AFM}_1({\Gamma}) =  S\sqrt{2K(3J + 2K)}]$.
In both models, the ground state and the magnon Hamiltonians are invariant under the spatial inversion
and the combined transformation of time-reversal and a spin rotation by 180$^\circ$ around the axis which is perpendicular to the ground state spin direction.
As a result, both models exhibit vanishing magnon Berry curvature across the entire Brillouin zone,
leading to the absence of the magnon Hall effect.

Despite the trivial magnon topology with vanishing magnon Berry curvature everywhere,
here we demonstrate that our models exhibit an intrinsic Hall effect of the SSC, namely the SSC Hall effect.
The SSC operator is defined as the counter-clockwise triple products around the upper and lower triangular unit cells:
${\hat \chi} = \hat {\v S}^{A}\cdot (\hat {\v S}^{B} \times \hat {\v S}^{C})$ and ${\hat {\bar \chi}} = \hat {\v S}^{A'}\cdot (\hat {\v S}^{B'} \times \hat {\v S}^{C})$
(see Fig.~\ref{fig:2}\textbf{a} and \textbf{e}).
Because the Hamiltonian is time-reversal symmetric and the SSC operator is invariant under cyclic permutations,
the equilibrium SSC profiles $\chi_{n,{\v k}}=\langle n|\hat \chi |n\rangle$ and $\bar\chi_{n,{\v k}}=\langle n|\hat {\bar{\chi}} |n\rangle$, where $|n\rangle$ is $n$-th eigenstate of magnons, are odd functions in the momentum space with $C_3$ rotational symmetry (see Supplementary Section I).
Consequently, their integrals over the Brillouin zone, {\it i.e.}, the SSCs of upper and lower triangles, vanish in equilibrium.
Additionally, it is important to note that $\bar \chi_{n,{\v k}} = \chi_{n,{-\v k}}$ due to the fact that the upper and lower triangles are connected by spatial inversion.
Combining this with $\chi_{n,{-\v k}} = -\chi_{n,{\v k}}$,
we deduce that the two spin chiralities are opposite, yielding $\bar \chi_{n,{\v k}} = -\chi_{n,{\v k}}$.
Therefore, when summing over the upper and lower triangles, the momentum profile of the net SSC, $\chi_{n,{\v k}}^{\rm net} \equiv \chi_{n,{\v k}} + \bar\chi_{n,{\v k}}$, cancels out and vanishes in equilibrium.

Although the net SSC profile vanishes in equilibrium, a nonequilibrium bias might generate intrinsic SSC Hall transport through the interband superpositions.
To check this possibility, we perform a heuristic calculation of the momentum-specific nonequilibrium SSC density induced by the temperature gradient along the $x$-direction using the Kubo response formalism as described in Ref.~\cite{Li2020a}.
Because $\chi_{n,{\v k}}^{\rm net} = 0$, the intraband contribution, representing the distribution shift effect of $\chi_{n,{\v k}}^{\rm net}$ vanishes.
However, the interband contribution of the nonequilibrium SSC from the two triangles is identical and does not cancel out.
In Fig.~\ref{fig:2}\textbf{c} and \textbf{g}, we present the momentum-resolved nonequilibrium SSC ($\delta \chi_{n,\v k}$) for the lowest magnonic state
(see Supplementary Section IV for calculation details and the nonequilibrium SSC profiles of all three magnonic states).
The opposite values of $\delta \chi_{n,\v k}$ between positive and negative $k_y$’s indicate that the nonequilibrium states with positive and negative SSC propagate along the transverse direction with opposite velocities,
hinting at the potential existence of the SSC Hall transport.

\begin{figure}[t]
\centering
\includegraphics[width=0.5\columnwidth]{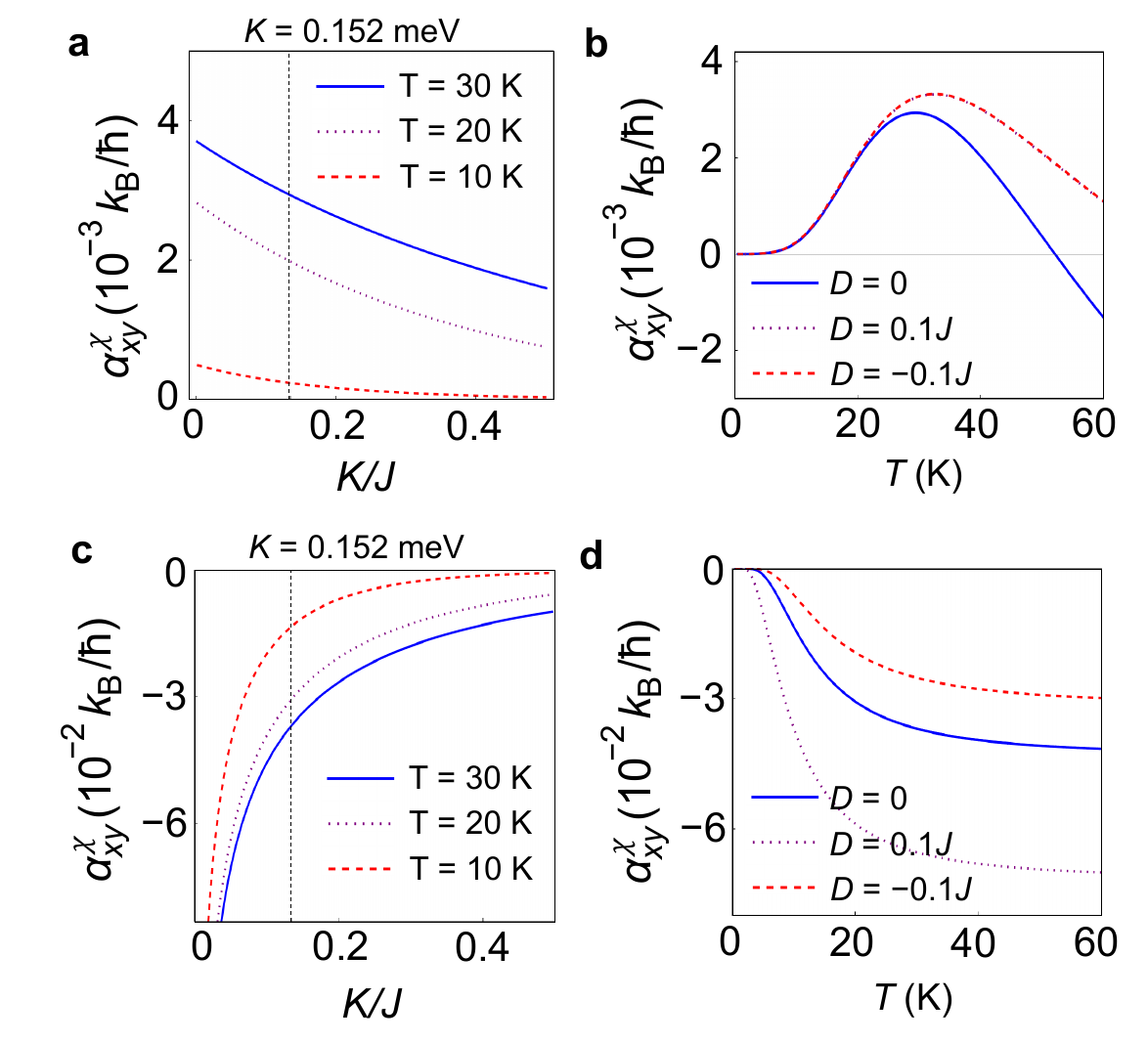}
\caption{\textbf{SSC Nernst conductivity.} \textbf{a}, Easy-axis anisotropy, and \textbf{b}, temperature, dependence of the SSC Nernst conductivity in the Kagome FM model.
\textbf{c}, Easy-axis anisotropy, and \textbf{d}, temperature, dependence of the SSC Nernst conductivity in the Kagome AFM model. The symbol $D$ in \textbf{b} and \textbf{d} is the out-of-plane DMI parameter (see Supplementary Section I). The dashed lines in \textbf{a} and \textbf{c} represent $K = 0.152$ meV.}\label{fig:3}
\end{figure}

To investigate the SSC Hall response rigorously by going beyond the aforementioned heuristic approach, we invoke the generalized Berry curvature method~\cite{Li2020} for the SSC, $\Omega^{\chi(\bar\chi)}_{n, \v k}$, which is directly related to the SSC Hall-type conductivities (Supplementary Section I).
Although the magnon Berry curvature is zero everywhere, the off-diagonal elements $\langle n|\hat {\chi} |m\rangle$ $(m\neq n)$ allow
the finite SSC Berry curvatures via interband superpositions.
The calculated SSC Berry curvatures exhibit a broad peak around the $\Gamma$-point,
where a dispersive band ($\epsilon_1$ or $\epsilon_2$) intersects with the flat band ($\epsilon_0$),
and relatively a narrower peak around the $K$-point, where the two dispersive bands converge (see Fig.~\ref{fig:2}\textbf{d} and \textbf{h}).
Additionally, we note that the SSC Berry curvatures of upper and lower triangles demonstrate identical behavior $\Omega^{\chi}_{n, \v k}  = \Omega^{\bar\chi}_{n, \v k}$. This indicates that the net SSC Hall transport does not vanish when averaging over the upper and lower triangles.

The generalized SSC Berry curvature induces a transverse current of the SSC in the presence of a temperature gradient $J^\chi_y = -\alpha^\chi_{xy} \partial_x T$~\cite{Li2020}.
Here, $\alpha^\chi_{xy} = \frac{k_B}{\hbar V} \sum_n \sum_\v k c_1(\rho_n) \Omega^{\chi}_{n, \v k}$ is the Nernst conductivity of the SSC,
where $k_B$ is the Boltzmann constant and $c_1(\rho) = (1+\rho){\rm ln}(1+\rho) - \rho {\rm ln}\rho$.
In Fig.~\ref{fig:3}, we illustrate the temperature dependence of the SSC Nernst conductivity.
This is our main result: the Kagome FM and AFM exhibit the SSC Hall transport even though the magnon Berry curvature is zero everywhere.
It is worth noting that in the coplanar AFM case, the SSC Nernst conductivity exhibits a relatively larger magnitude compared to that in the collinear FM case.
This contrast arises because the dominant contribution of the SSC Berry curvature emerges near the flat band~(Fig.~\ref{fig:2}\textbf{d} and \textbf{h}).
In the collinear FM model, the flat band energy is at the top of the bands, whereas in the coplanar AFM model, it is at the bottom of the bands~(Fig.~\ref{fig:2}\textbf{b} and \textbf{f}).
As a result, in the coplanar AFM model, the magnitude of the SSC Hall conductivity drastically increases with decreasing $K$~(Fig.~\ref{fig:3}\textbf{c} and \textbf{d}).
Additionally, we emphasize that the SSC Hall conductivity remains finite for $K\rightarrow 0$ limit.
This indicates that the SSC Hall effect is a distinctive feature of the Kagome magnetic materials,
stemming exclusively from the exchange interaction and the lattice geometry.

In the Supplementary Information, we also calculate the DMI effect on the SSC Nernst conductivity in both models.
For the Kagome FM, the influence of the DMI on the SSC Nernst conductivity can be summarized as follows.
At low temperatures, where the contribution is primarily from the lowest band, the SSC Nernst conductivity is minimally influenced by the DMI coefficient. However, as the temperature increases and the contribution from the middle band becomes dominant, the conductivity undergoes a sign reversal, with its magnitude increasing as the magnitude of the DMI coefficient decreases. The sign of the DMI coefficient does not affect this behavior from the perspective of symmetry.
However, for the Kagome AFM, the SSC Nernst conductivity either increases or decreases depending on the sign of the DMI coefficient,
due to the DMI's influence on the flat band position (see Supplementary Section I).

\begin{figure}[t]
\centering
\includegraphics[width=0.5\columnwidth]{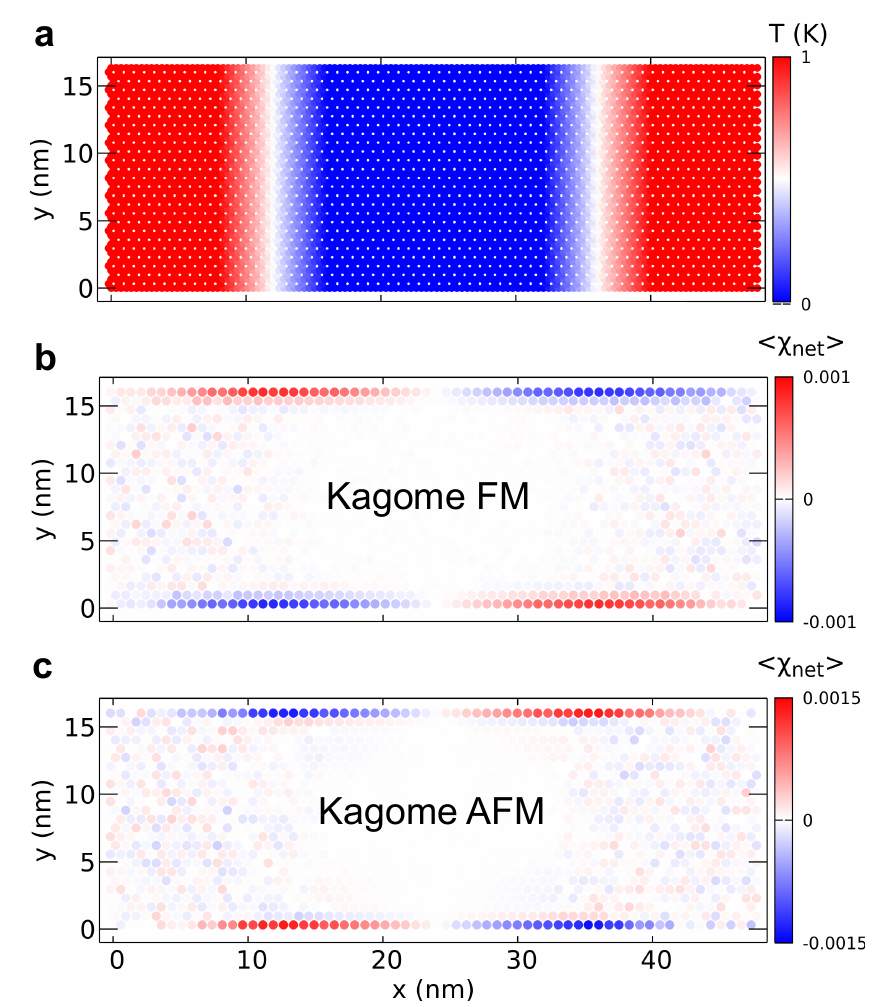}
\caption{\textbf{Atomistic simulation results.} \textbf{a} Temperature profile. \textbf{b} and \textbf{c} net SSC accumulation profiles for Kagome FM and AFM, respectively.
The red and blue dots in \textbf{b} and \textbf{c} represent positive and negative net SSC, respectively.}\label{fig:4}
\end{figure}

\section{Scalar spin chirality accumulations}

To confirm our theoretical prediction of the SSC Hall effect, we perform atomistic spin dynamics simulations by solving the stochastic Landau-Lifshitz-Gilbert equation (Method).
Our simulation setup involves a rectangular stripe of a Kagome lattice with a finite width in the $y$-direction and periodic boundary conditions along the longer $x$-direction. The simulations are conducted with a nonuniform temperature profile shown in  Figs.~\ref{fig:4}\textbf{a}. The numerical results of the net SSC accumulation for Kagome FMs and AFMs are shown in Figs.~\ref{fig:4}\textbf{b} and \textbf{c}, respectively, where the red and blue dots represent positive and negative values of the net SSC.
We observe that the opposite SSC accumulation occurs at both edges of the sample in regions with a finite temperature gradient.
Note that Kagome FM and AFM models show the opposite signs of the SSC edge accumulations. This sign difference of the two models agrees with our theoretical results that the two models possess the opposite signs of the SSC Nernst conductivities as shown in Fig.~\ref{fig:3}.
Our numerical simulation confirms that a temperature gradient along the $x$-direction induces the SSC accumulation due to the transverse SSC current flowing along the $y$-direction.

For a rough quantitative estimation of the SSC accumulation at finite temperatures in bulk Kagome magnets, we use the solution of the drift-diffusion equation for the SSC,
$\chi(y)=\alpha^\chi_{xy}\frac{\tau\sinh\left(\frac{W-2y}{2\lambda}\right)}{\lambda\cosh\left(\frac{W}{2\lambda}\right)}\partial_x T$,
where $W$ is the width of the sample, $\lambda$ and $\tau$ are the SSC diffusion length and relaxation time, respectively.
For chosen parameters: $\alpha^\chi_{xy} = 10^{-2}$ $k_B/\hbar$, $W = 20$ nm, $\partial_x T = 10$ K/$\mu \rm m$, $\lambda = 1$ nm and $\tau = 0.1$ ns. The resulting SSC accumulation at the edge is $\chi \sim 1\times10^{-3}$/cell (see Supplementary Section V for SSC profiles by using other values of $\lambda$ and $\tau$).

\section{Discussion}

In this work, we have predicted the SSC Hall effect in Kagome FM and AFM models.
Going beyond the previous studies on magnonic Hall-like phenomena that require the finite magnon Berry curvature, we have found that the SSC Hall effect exists in Kagome magnets in the complete absence of the magnon Berry curvature.
This suggests that the SSC Hall effect is an intrinsic property of Kagome magnets and does not depend on specific symmetry breaking or spin-orbit coupling effects.
Additionally, we have confirmed the SSC Hall effect by numerical simulations of the finite-temperature spin dynamics in Kagome magnets.
We note that while we have used a temperature gradient to drive SSC transport,
an alternative method using electronic techniques such as the spin Hall effect might be used to pump the SSC~\cite{Cornelissen2015,Ochoa2016}.

For the potential experimental schemes for detecting the magnonic SSC, we examine the electron orbital magnetization induced by the SSC.
It is known that the SSC induces the so-called topological orbital moment~\cite{Hoffmann2015b, Dias2016, Hanke2016a, Lux2018a, Grytsiuk2020, Zhang2020a}, which is given phenomenologically by $L = \kappa^{\rm TO} \chi$ with $\kappa^{\rm TO}$ the topological orbital susceptibility.
To compute this, we utilize a simple tight-binding model as described in Ref.~\cite{Zhang2020a},
and apply the modern theory of electron orbital magnetization~\cite{Thonhauser2005,Shi2007,Xiao2010b}
in the presence of a noncollinear spin texture due to the SSC.
For $\kappa^{\rm TO}$ = 0.1 $\mu_B$ with parameters used in previous section,
the maximum value of the topological orbital magnetization is estimated to be on the order of $10^{-4} \mu_B/{\rm cell}$ (Supplementary Section VI), which is comparable
to the orbital accumulation in a light metal Ti that has been observed successfully by using magneto-optical Kerr effect~\cite{Choi2023}.
Additionally, we would like to remark that it is plausible to consider that the accumulation of the nonequilibrium SSC at the edge can induce an electromotive force in a proximate normal metal~\cite{Ochoa2016}, making it detectable through an electrical measurement scheme.
Furthermore, the spin chiralities in bulk materials can be measured by neutron scattering experiments~\cite{Maleyev1995a, Maleyev1998,Simonet2012a,Lee2013} and the resonant inelastic x-ray scattering experiments~\cite{Ko2011,Xiong2020}.

For the lattice systems other than the Kagome ones, we verified that the SSC Hall effect is also present in honeycomb lattice magnets with vanishing magnon Berry curvature (Supplementary Section I and III), but is absent in triangular lattice magnets, which calls for the need to develop a classification scheme for the SSC Hall effect based on the lattice structure. By lifting the common assumption that the SSC is a passive background in this work, we have uncovered one transport phenomenon, the SSC Hall effect. We envision that our finding will trigger further investigations of the diverse transport phenomena of the SSC and thereby lead to a more comprehensive understanding of chiral phenomena where both the active and the passive roles of the SSC are treated on equal footing.

\section{Method}

For the atomistic spin simulation, we use the stochastic Landau-Lifshitz-Gilbert equation to study the temperature dependent magnetic dynamics, which can be expressed as
\begin{equation}
\frac{\partial{\v S}_i}{\partial t}~=~-\gamma({\v S}_i \times {\v H}_i^{\rm eff})+\alpha\left({\v S}_i \times \frac{\partial{\v S}_i}{\partial t}\right)
\end{equation}
where $\gamma$ is the gyromagnetic ratio, $\alpha$ is the intrinsic Gilbert Damping and the site spin ${\v S}_i$ is a normalized spin vector. The effective field of each spin, ${\v H}_i^{\rm eff}~=~-\frac{1}{\mu_S}\frac{\partial H}{\partial {\v S}_i}~+~{\v H}_i^{\rm th}$, where $\mu_S$ is the atomic magnetic moment and ${\v H}_i^{\rm th}$ is the spatial dependent thermal field. The temperature effects of each spin is described by the Langevin dynamics and is expressed as
${\v H}_i^{\rm th}~=~\Gamma(t)\sqrt{\frac{2\alpha k_B T}{\gamma \mu_S \Delta t}}$, where $\Gamma(t)$ is the three dimensional
Gaussian distribution with a mean zero, $k_B$ is the Boltzmann constant and $\Delta t$ is the integration time step \cite{Brown1979,Evans2014}.

To study the spatial temperature dependence on the spin system, we integrate the  stochastic Landau-Lifshitz-Gilbert equation using the Heun method.
For the numerical simulations, we take the parameters $\alpha$ = 0.01, $\Delta t$ = 5 fs.
The simulations are performed on Kagome lattices, with a periodic boundary condition along the $x$-direction to minimize thermal noise at the edges, while the $y$-direction is open to observe the magnon effect at the edges.
The spatially dependent temperature profile of the system is shown in Fig.~\ref{fig:4}\textbf{a}, with minimum and maximum temperatures of 0.01 K and 1 K, respectively.
Initially, the spins are thermalized for $3\times 10^6$ time steps and the ensemble averages of scalar spin chiralities are performed
for $2 \times 10^8$ time steps.

\section*{Acknowledgement}

We thank Kouki Nakata for useful discussions.
G.G. acknowledges support by the National Research Foundation of Korea (NRF-2022R1C1C2006578).
S.K.K. was supported by the Brain Pool Plus Program through the National Research Foundation of Korea funded by the Ministry of Science and ICT (2020H1D3A2A03099291) and Samsung Science and Technology Foundation (SSTF-BA2202-04).
Y.T. is supported by NSF under Grant No. DMR-2049979.

\bibliography{reference_SC}

\begin{thebibliography}{10}
\expandafter\ifx\csname url\endcsname\relax
  \def\url#1{\burl{#1}}\fi
\expandafter\ifx\csname urlprefix\endcsname\relax\def\urlprefix{URL }\fi
\providecommand{\bibinfo}[2]{#2}
\providecommand{\eprint}[2][]{\url{#2}}
\providecommand{\doi}[1]{\url{https://doi.org/#1}}
\bibcommenthead

\bibitem{Wen1989}
\bibinfo{author}{Wen, X.~G.}, \bibinfo{author}{Wilczek, F.} \&
  \bibinfo{author}{Zee, A.}
\newblock \bibinfo{title}{Chiral spin states and superconductivity}.
\newblock \emph{\bibinfo{journal}{Phys. Rev. B}} \textbf{\bibinfo{volume}{39}},
  \bibinfo{pages}{11413--11423} (\bibinfo{year}{1989}).

\bibitem{Hoffmann2015b}
\bibinfo{author}{Hoffmann, M.} \emph{et~al.}
\newblock \bibinfo{title}{Topological orbital magnetization and emergent
  {{Hall}} effect of an atomic-scale spin lattice at a surface}.
\newblock \emph{\bibinfo{journal}{Phys. Rev. B}} \textbf{\bibinfo{volume}{92}},
  \bibinfo{pages}{020401} (\bibinfo{year}{2015}).

\bibitem{Dias2016}
\bibinfo{author}{{dos Santos Dias}, M.}, \bibinfo{author}{Bouaziz, J.},
  \bibinfo{author}{Bouhassoune, M.}, \bibinfo{author}{Bl{\"u}gel, S.} \&
  \bibinfo{author}{Lounis, S.}
\newblock \bibinfo{title}{Chirality-driven orbital magnetic moments as a new
  probe for topological magnetic structures}.
\newblock \emph{\bibinfo{journal}{Nat. Commun.}} \textbf{\bibinfo{volume}{7}},
  \bibinfo{pages}{13613} (\bibinfo{year}{2016}).

\bibitem{Hanke2016a}
\bibinfo{author}{Hanke, J.-P.} \emph{et~al.}
\newblock \bibinfo{title}{Role of {{Berry}} phase theory for describing orbital
  magnetism: {{From}} magnetic heterostructures to topological orbital
  ferromagnets}.
\newblock \emph{\bibinfo{journal}{Phys. Rev. B}} \textbf{\bibinfo{volume}{94}},
  \bibinfo{pages}{121114} (\bibinfo{year}{2016}).

\bibitem{Lux2018a}
\bibinfo{author}{Lux, F.~R.}, \bibinfo{author}{Freimuth, F.},
  \bibinfo{author}{Bl{\"u}gel, S.} \& \bibinfo{author}{Mokrousov, Y.}
\newblock \bibinfo{title}{Engineering chiral and topological orbital magnetism
  of domain walls and skyrmions}.
\newblock \emph{\bibinfo{journal}{Commun. Phys.}} \textbf{\bibinfo{volume}{1}},
  \bibinfo{pages}{1--8} (\bibinfo{year}{2018}).

\bibitem{Grytsiuk2020}
\bibinfo{author}{Grytsiuk, S.} \emph{et~al.}
\newblock \bibinfo{title}{Topological--chiral magnetic interactions driven by
  emergent orbital magnetism}.
\newblock \emph{\bibinfo{journal}{Nat. Commun.}} \textbf{\bibinfo{volume}{11}},
  \bibinfo{pages}{511} (\bibinfo{year}{2020}).

\bibitem{Zhang2020a}
\bibinfo{author}{Zhang, L.-c.} \emph{et~al.}
\newblock \bibinfo{title}{Imprinting and driving electronic orbital magnetism
  using magnons}.
\newblock \emph{\bibinfo{journal}{Commun Phys}} \textbf{\bibinfo{volume}{3}},
  \bibinfo{pages}{1--8} (\bibinfo{year}{2020}).

\bibitem{Taguchi2001}
\bibinfo{author}{Taguchi, Y.}, \bibinfo{author}{Oohara, Y.},
  \bibinfo{author}{Yoshizawa, H.}, \bibinfo{author}{Nagaosa, N.} \&
  \bibinfo{author}{Tokura, Y.}
\newblock \bibinfo{title}{Spin {{Chirality}}, {{Berry Phase}}, and {{Anomalous
  Hall Effect}} in a {{Frustrated Ferromagnet}}}.
\newblock \emph{\bibinfo{journal}{Science}} \textbf{\bibinfo{volume}{291}},
  \bibinfo{pages}{2573--2576} (\bibinfo{year}{2001}).

\bibitem{Tatara2002}
\bibinfo{author}{Tatara, G.} \& \bibinfo{author}{Kawamura, H.}
\newblock \bibinfo{title}{Chirality-{{Driven Anomalous Hall Effect}} in {{Weak
  Coupling Regime}}}.
\newblock \emph{\bibinfo{journal}{J. Phys. Soc. Jpn.}}
  \textbf{\bibinfo{volume}{71}}, \bibinfo{pages}{2613--2616}
  (\bibinfo{year}{2002}).

\bibitem{Machida2010}
\bibinfo{author}{Machida, Y.}, \bibinfo{author}{Nakatsuji, S.},
  \bibinfo{author}{Onoda, S.}, \bibinfo{author}{Tayama, T.} \&
  \bibinfo{author}{Sakakibara, T.}
\newblock \bibinfo{title}{Time-reversal symmetry breaking and spontaneous
  {{Hall}} effect without magnetic dipole order}.
\newblock \emph{\bibinfo{journal}{Nature}} \textbf{\bibinfo{volume}{463}},
  \bibinfo{pages}{210--213} (\bibinfo{year}{2010}).

\bibitem{Ishizuka2018}
\bibinfo{author}{Ishizuka, H.} \& \bibinfo{author}{Nagaosa, N.}
\newblock \bibinfo{title}{Spin chirality induced skew scattering and anomalous
  {{Hall}} effect in chiral magnets}.
\newblock \emph{\bibinfo{journal}{Sci. Adv.}} \textbf{\bibinfo{volume}{4}},
  \bibinfo{pages}{eaap9962} (\bibinfo{year}{2018}).

\bibitem{Katsura2010}
\bibinfo{author}{Katsura, H.}, \bibinfo{author}{Nagaosa, N.} \&
  \bibinfo{author}{Lee, P.~A.}
\newblock \bibinfo{title}{Theory of the {{Thermal Hall Effect}} in {{Quantum
  Magnets}}}.
\newblock \emph{\bibinfo{journal}{Phys. Rev. Lett.}}
  \textbf{\bibinfo{volume}{104}}, \bibinfo{pages}{066403}
  (\bibinfo{year}{2010}).

\bibitem{Oh2024}
\bibinfo{author}{Oh, T.} \& \bibinfo{author}{Nagaosa, N.}
\newblock \bibinfo{title}{Phonon thermal {{Hall}} effect in {{Mott}} insulators
  via skew-scattering by the scalar spin chirality.}
  \bibinfo{pages}{arXiv:2408.01671}.

\bibitem{Shindou2001}
\bibinfo{author}{Shindou, R.} \& \bibinfo{author}{Nagaosa, N.}
\newblock \bibinfo{title}{Orbital {{Ferromagnetism}} and {{Anomalous Hall
  Effect}} in {{Antiferromagnets}} on the {{Distorted}} fcc {{Lattice}}}.
\newblock \emph{\bibinfo{journal}{Phys. Rev. Lett.}}
  \textbf{\bibinfo{volume}{87}}, \bibinfo{pages}{116801}
  (\bibinfo{year}{2001}).

\bibitem{Nakatsuji2015}
\bibinfo{author}{Nakatsuji, S.}, \bibinfo{author}{Kiyohara, N.} \&
  \bibinfo{author}{Higo, T.}
\newblock \bibinfo{title}{Large anomalous {{Hall}} effect in a non-collinear
  antiferromagnet at room temperature}.
\newblock \emph{\bibinfo{journal}{Nature}} \textbf{\bibinfo{volume}{527}},
  \bibinfo{pages}{212--215} (\bibinfo{year}{2015}).

\bibitem{Han2017}
\bibinfo{author}{Han, J.~H.} \& \bibinfo{author}{Lee, H.}
\newblock \bibinfo{title}{Spin {{Chirality}} and {{Hall-Like Transport
  Phenomena}} of {{Spin Excitations}}}.
\newblock \emph{\bibinfo{journal}{J. Phys. Soc. Jpn.}}
  \textbf{\bibinfo{volume}{86}}, \bibinfo{pages}{011007}
  (\bibinfo{year}{2017}).

\bibitem{Kim2024}
\bibinfo{author}{Kim, H.-L.} \emph{et~al.}
\newblock \bibinfo{title}{Thermal {{Hall}} effects due to topological spin
  fluctuations in {{YMnO$_3$}}}.
\newblock \emph{\bibinfo{journal}{Nat Commun}} \textbf{\bibinfo{volume}{15}},
  \bibinfo{pages}{243} (\bibinfo{year}{2024}).

\bibitem{Neubauer2009}
\bibinfo{author}{Neubauer, A.} \emph{et~al.}
\newblock \bibinfo{title}{Topological {{Hall Effect}} in the {{A Phase}} of
  {{MnSi}}}.
\newblock \emph{\bibinfo{journal}{Phys. Rev. Lett.}}
  \textbf{\bibinfo{volume}{102}}, \bibinfo{pages}{186602}
  (\bibinfo{year}{2009}).

\bibitem{Kanazawa2011}
\bibinfo{author}{Kanazawa, N.} \emph{et~al.}
\newblock \bibinfo{title}{Large {{Topological Hall Effect}} in a {{Short-Period
  Helimagnet MnGe}}}.
\newblock \emph{\bibinfo{journal}{Phys. Rev. Lett.}}
  \textbf{\bibinfo{volume}{106}}, \bibinfo{pages}{156603}
  (\bibinfo{year}{2011}).

\bibitem{Franz2014}
\bibinfo{author}{Franz, C.} \emph{et~al.}
\newblock \bibinfo{title}{Real-{{Space}} and {{Reciprocal-Space Berry Phases}}
  in the {{Hall Effect}} of {{Mn}}$_{1-x}${{Fe}}$_x${{Si}}}.
\newblock \emph{\bibinfo{journal}{Phys. Rev. Lett.}}
  \textbf{\bibinfo{volume}{112}}, \bibinfo{pages}{186601}
  (\bibinfo{year}{2014}).

\bibitem{vanHoogdalem2013}
\bibinfo{author}{{van Hoogdalem}, K.~A.}, \bibinfo{author}{Tserkovnyak, Y.} \&
  \bibinfo{author}{Loss, D.}
\newblock \bibinfo{title}{Magnetic texture-induced thermal {{Hall}} effects}.
\newblock \emph{\bibinfo{journal}{Phys. Rev. B}} \textbf{\bibinfo{volume}{87}},
  \bibinfo{pages}{024402} (\bibinfo{year}{2013}).

\bibitem{Kim2019b}
\bibinfo{author}{Kim, S.~K.}, \bibinfo{author}{Nakata, K.},
  \bibinfo{author}{Loss, D.} \& \bibinfo{author}{Tserkovnyak, Y.}
\newblock \bibinfo{title}{Tunable {{Magnonic Thermal Hall Effect}} in
  {{Skyrmion Crystal Phases}} of {{Ferrimagnets}}}.
\newblock \emph{\bibinfo{journal}{Phys. Rev. Lett.}}
  \textbf{\bibinfo{volume}{122}}, \bibinfo{pages}{057204}
  (\bibinfo{year}{2019}).

\bibitem{Xu2024}
\bibinfo{author}{Xu, S.} \emph{et~al.}
\newblock \bibinfo{title}{Universal scaling law for chiral antiferromagnetism}.
\newblock \emph{\bibinfo{journal}{Nat Commun}} \textbf{\bibinfo{volume}{15}},
  \bibinfo{pages}{3717} (\bibinfo{year}{2024}).

\bibitem{Onose2010}
\bibinfo{author}{Onose, Y.} \emph{et~al.}
\newblock \bibinfo{title}{Observation of the {{Magnon Hall Effect}}}.
\newblock \emph{\bibinfo{journal}{Science}} \textbf{\bibinfo{volume}{329}},
  \bibinfo{pages}{297--299} (\bibinfo{year}{2010}).

\bibitem{Matsumoto2011}
\bibinfo{author}{Matsumoto, R.} \& \bibinfo{author}{Murakami, S.}
\newblock \bibinfo{title}{Theoretical {{Prediction}} of a {{Rotating Magnon
  Wave Packet}} in {{Ferromagnets}}}.
\newblock \emph{\bibinfo{journal}{Phys. Rev. Lett.}}
  \textbf{\bibinfo{volume}{106}}, \bibinfo{pages}{197202}
  (\bibinfo{year}{2011}).

\bibitem{Matsumoto2014a}
\bibinfo{author}{Matsumoto, R.}, \bibinfo{author}{Shindou, R.} \&
  \bibinfo{author}{Murakami, S.}
\newblock \bibinfo{title}{Thermal {{Hall}} effect of magnons in magnets with
  dipolar interaction}.
\newblock \emph{\bibinfo{journal}{Phys. Rev. B}} \textbf{\bibinfo{volume}{89}},
  \bibinfo{pages}{054420} (\bibinfo{year}{2014}).

\bibitem{Mook2014a}
\bibinfo{author}{Mook, A.}, \bibinfo{author}{Henk, J.} \&
  \bibinfo{author}{Mertig, I.}
\newblock \bibinfo{title}{Magnon {{Hall}} effect and topology in kagome
  lattices: {{A}} theoretical investigation}.
\newblock \emph{\bibinfo{journal}{Phys. Rev. B}} \textbf{\bibinfo{volume}{89}},
  \bibinfo{pages}{134409} (\bibinfo{year}{2014}).

\bibitem{Kim2016a}
\bibinfo{author}{Kim, S.~K.}, \bibinfo{author}{Ochoa, H.},
  \bibinfo{author}{Zarzuela, R.} \& \bibinfo{author}{Tserkovnyak, Y.}
\newblock \bibinfo{title}{Realization of the {{Haldane-Kane-Mele Model}} in a
  {{System}} of {{Localized Spins}}}.
\newblock \emph{\bibinfo{journal}{Phys. Rev. Lett.}}
  \textbf{\bibinfo{volume}{117}}, \bibinfo{pages}{227201}
  (\bibinfo{year}{2016}).

\bibitem{Owerre2016}
\bibinfo{author}{Owerre, S.~A.}
\newblock \bibinfo{title}{A first theoretical realization of honeycomb
  topological magnon insulator}.
\newblock \emph{\bibinfo{journal}{J. Phys.: Condens. Matter}}
  \textbf{\bibinfo{volume}{28}}, \bibinfo{pages}{386001}
  (\bibinfo{year}{2016}).

\bibitem{Cheng2016}
\bibinfo{author}{Cheng, R.}, \bibinfo{author}{Okamoto, S.} \&
  \bibinfo{author}{Xiao, D.}
\newblock \bibinfo{title}{Spin {{Nernst Effect}} of {{Magnons}} in {{Collinear
  Antiferromagnets}}}.
\newblock \emph{\bibinfo{journal}{Phys. Rev. Lett.}}
  \textbf{\bibinfo{volume}{117}}, \bibinfo{pages}{217202}
  (\bibinfo{year}{2016}).

\bibitem{Zyuzin2016}
\bibinfo{author}{Zyuzin, V.~A.} \& \bibinfo{author}{Kovalev, A.~A.}
\newblock \bibinfo{title}{Magnon {{Spin Nernst Effect}} in
  {{Antiferromagnets}}}.
\newblock \emph{\bibinfo{journal}{Phys. Rev. Lett.}}
  \textbf{\bibinfo{volume}{117}}, \bibinfo{pages}{217203}
  (\bibinfo{year}{2016}).

\bibitem{Shiomi2017}
\bibinfo{author}{Shiomi, Y.}, \bibinfo{author}{Takashima, R.} \&
  \bibinfo{author}{Saitoh, E.}
\newblock \bibinfo{title}{Experimental evidence consistent with a magnon
  {{Nernst}} effect in the antiferromagnetic insulator {{MnPS}}$_3$}.
\newblock \emph{\bibinfo{journal}{Phys. Rev. B}} \textbf{\bibinfo{volume}{96}},
  \bibinfo{pages}{134425} (\bibinfo{year}{2017}).

\bibitem{Zhang2022}
\bibinfo{author}{Zhang, H.} \& \bibinfo{author}{Cheng, R.}
\newblock \bibinfo{title}{A perspective on magnon spin {{Nernst}} effect in
  antiferromagnets}.
\newblock \emph{\bibinfo{journal}{Appl. Phys. Lett.}}
  \textbf{\bibinfo{volume}{120}}, \bibinfo{pages}{090502}
  (\bibinfo{year}{2022}).

\bibitem{Go2024b}
\bibinfo{author}{Go, G.}, \bibinfo{author}{An, D.}, \bibinfo{author}{Lee,
  H.-W.} \& \bibinfo{author}{Kim, S.~K.}
\newblock \bibinfo{title}{Magnon {{Orbital Nernst Effect}} in {{Honeycomb
  Antiferromagnets}} without {{Spin}}--{{Orbit Coupling}}}.
\newblock \emph{\bibinfo{journal}{Nano Lett.}} \textbf{\bibinfo{volume}{24}},
  \bibinfo{pages}{5968--5974} (\bibinfo{year}{2024}).

\bibitem{You2023}
\bibinfo{author}{You, J.~Y.} \& \bibinfo{author}{Feng, Y.~P.}
\newblock \bibinfo{title}{A two-dimensional kagome magnet with tunable
  topological phases}.
\newblock \emph{\bibinfo{journal}{Mater. Today Chem.}}
  \textbf{\bibinfo{volume}{30}}, \bibinfo{pages}{101566}
  (\bibinfo{year}{2023}).

\bibitem{Li2020a}
\bibinfo{author}{Li, B.}, \bibinfo{author}{Mook, A.},
  \bibinfo{author}{Raeliarijaona, A.} \& \bibinfo{author}{Kovalev, A.~A.}
\newblock \bibinfo{title}{Magnonic analog of the {{Edelstein}} effect in
  antiferromagnetic insulators}.
\newblock \emph{\bibinfo{journal}{Phys. Rev. B}}
  \textbf{\bibinfo{volume}{101}}, \bibinfo{pages}{024427}
  (\bibinfo{year}{2020}).

\bibitem{Li2020}
\bibinfo{author}{Li, B.}, \bibinfo{author}{Sandhoefner, S.} \&
  \bibinfo{author}{Kovalev, A.~A.}
\newblock \bibinfo{title}{Intrinsic spin {{Nernst}} effect of magnons in a
  noncollinear antiferromagnet}.
\newblock \emph{\bibinfo{journal}{Phys. Rev. Res.}}
  \textbf{\bibinfo{volume}{2}}, \bibinfo{pages}{013079} (\bibinfo{year}{2020}).

\bibitem{Cornelissen2015}
\bibinfo{author}{Cornelissen, L.~J.}, \bibinfo{author}{Liu, J.},
  \bibinfo{author}{Duine, R.~A.}, \bibinfo{author}{Youssef, J.~B.} \&
  \bibinfo{author}{{van Wees}, B.~J.}
\newblock \bibinfo{title}{Long-distance transport of magnon spin information in
  a magnetic insulator at room temperature}.
\newblock \emph{\bibinfo{journal}{Nat. Phys.}} \textbf{\bibinfo{volume}{11}},
  \bibinfo{pages}{1022--1026} (\bibinfo{year}{2015}).

\bibitem{Ochoa2016}
\bibinfo{author}{Ochoa, H.}, \bibinfo{author}{Kim, S.~K.} \&
  \bibinfo{author}{Tserkovnyak, Y.}
\newblock \bibinfo{title}{Topological spin-transfer drag driven by skyrmion
  diffusion}.
\newblock \emph{\bibinfo{journal}{Phys. Rev. B}} \textbf{\bibinfo{volume}{94}},
  \bibinfo{pages}{024431} (\bibinfo{year}{2016}).

\bibitem{Thonhauser2005}
\bibinfo{author}{Thonhauser, T.}, \bibinfo{author}{Ceresoli, D.},
  \bibinfo{author}{Vanderbilt, D.} \& \bibinfo{author}{Resta, R.}
\newblock \bibinfo{title}{Orbital {{Magnetization}} in {{Periodic
  Insulators}}}.
\newblock \emph{\bibinfo{journal}{Phys. Rev. Lett.}}
  \textbf{\bibinfo{volume}{95}}, \bibinfo{pages}{137205}
  (\bibinfo{year}{2005}).

\bibitem{Shi2007}
\bibinfo{author}{Shi, J.}, \bibinfo{author}{Vignale, G.},
  \bibinfo{author}{Xiao, D.} \& \bibinfo{author}{Niu, Q.}
\newblock \bibinfo{title}{Quantum {{Theory}} of {{Orbital Magnetization}} and
  {{Its Generalization}} to {{Interacting Systems}}}.
\newblock \emph{\bibinfo{journal}{Phys. Rev. Lett.}}
  \textbf{\bibinfo{volume}{99}}, \bibinfo{pages}{197202}
  (\bibinfo{year}{2007}).

\bibitem{Xiao2010b}
\bibinfo{author}{Xiao, D.}, \bibinfo{author}{Chang, M.-C.} \&
  \bibinfo{author}{Niu, Q.}
\newblock \bibinfo{title}{Berry phase effects on electronic properties}.
\newblock \emph{\bibinfo{journal}{Rev. Mod. Phys.}}
  \textbf{\bibinfo{volume}{82}}, \bibinfo{pages}{1959--2007}
  (\bibinfo{year}{2010}).

\bibitem{Choi2023}
\bibinfo{author}{Choi, Y.-G.} \emph{et~al.}
\newblock \bibinfo{title}{Observation of the orbital {{Hall}} effect in a light
  metal {{Ti}}}.
\newblock \emph{\bibinfo{journal}{Nature}} \textbf{\bibinfo{volume}{619}},
  \bibinfo{pages}{52--56} (\bibinfo{year}{2023}).

\bibitem{Maleyev1995a}
\bibinfo{author}{Maleyev, S.~V.}
\newblock \bibinfo{title}{Investigation of {{Spin Chirality}} by {{Polarized
  Neutrons}}}.
\newblock \emph{\bibinfo{journal}{Phys. Rev. Lett.}}
  \textbf{\bibinfo{volume}{75}}, \bibinfo{pages}{4682--4685}
  (\bibinfo{year}{1995}).

\bibitem{Maleyev1998}
\bibinfo{author}{Maleyev, S.~V.} \emph{et~al.}
\newblock \bibinfo{title}{The first observation of dynamical chirality by means
  of polarized neutron scattering in the triangular-lattice antiferromagnet}.
\newblock \emph{\bibinfo{journal}{J. Phys.: Condens. Matter}}
  \textbf{\bibinfo{volume}{10}}, \bibinfo{pages}{951} (\bibinfo{year}{1998}).

\bibitem{Simonet2012a}
\bibinfo{author}{Simonet, V.}, \bibinfo{author}{Loire, M.} \&
  \bibinfo{author}{Ballou, R.}
\newblock \bibinfo{title}{Magnetic chirality as probed by neutron scattering}.
\newblock \emph{\bibinfo{journal}{Eur. Phys. J. Spec. Top.}}
  \textbf{\bibinfo{volume}{213}}, \bibinfo{pages}{5--36}
  (\bibinfo{year}{2012}).

\bibitem{Lee2013}
\bibinfo{author}{Lee, P.~A.} \& \bibinfo{author}{Nagaosa, N.}
\newblock \bibinfo{title}{Proposal to use neutron scattering to access scalar
  spin chirality fluctuations in kagome lattices}.
\newblock \emph{\bibinfo{journal}{Phys. Rev. B}} \textbf{\bibinfo{volume}{87}},
  \bibinfo{pages}{064423} (\bibinfo{year}{2013}).

\bibitem{Ko2011}
\bibinfo{author}{Ko, W.-H.} \& \bibinfo{author}{Lee, P.~A.}
\newblock \bibinfo{title}{Proposal for detecting spin-chirality terms in
  {{Mott}} insulators via resonant inelastic x-ray scattering}.
\newblock \emph{\bibinfo{journal}{Phys. Rev. B}} \textbf{\bibinfo{volume}{84}},
  \bibinfo{pages}{125102} (\bibinfo{year}{2011}).

\bibitem{Xiong2020}
\bibinfo{author}{Xiong, Z.}, \bibinfo{author}{Datta, T.} \&
  \bibinfo{author}{Yao, D.-X.}
\newblock \bibinfo{title}{Resonant inelastic x-ray scattering study of vector
  chiral ordered kagome antiferromagnet}.
\newblock \emph{\bibinfo{journal}{npj Quantum Mater.}}
  \textbf{\bibinfo{volume}{5}}, \bibinfo{pages}{1--9} (\bibinfo{year}{2020}).

\bibitem{Brown1979}
\bibinfo{author}{Brown, W.}
\newblock \bibinfo{title}{Thermal fluctuation of fine ferromagnetic particles}.
\newblock \emph{\bibinfo{journal}{IEEE Transactions on Magnetics}}
  \textbf{\bibinfo{volume}{15}}, \bibinfo{pages}{1196--1208}
  (\bibinfo{year}{1979}).

\bibitem{Evans2014}
\bibinfo{author}{Evans, R. F.~L.} \emph{et~al.}
\newblock \bibinfo{title}{Atomistic spin model simulations of magnetic
  nanomaterials}.
\newblock \emph{\bibinfo{journal}{J. Phys.: Condens. Matter}}
  \textbf{\bibinfo{volume}{26}}, \bibinfo{pages}{103202}
  (\bibinfo{year}{2014}).

\end{thebibliography}

\end{document}